\documentclass[aps,a4paper,showpacs,showkeys]{article}
\usepackage[twoside,top=2cm,bottom=2cm,left=2cm,right=2cm]{geometry}
\usepackage[utf8]{inputenc}	%lettere accentate da tastiera
\usepackage{authblk} % Afiliazione autori
\usepackage{epsfig}
\usepackage{amsmath}
\usepackage{amsfonts}
\usepackage{amssymb}
\usepackage{graphicx}
\usepackage{colordvi}
\usepackage[title]{appendix}
\usepackage{soul,color}
\date{}
\begin{document}
	\author{ Filippo Maimone$^{(1)}$\footnote{%
			e-mail address: \textit{filippo.maimone@gmail.com}}, Giovanni Scelza$^{(1)}$\footnote{%
			e-mail address: \textit{lucasce73@gmail.com}}, Adele Naddeo$^{(2)}$\footnote{%
			e-mail address: \textit{anaddeo@na.infn.it}}}
	
\affil{\small{\textit{$^{(1)}$ Associazione Culturale ``Velia Polis", via Capo di Mezzo, $84078$ Vallo della Lucania, Italy}}}
\affil{\small{\textit{$^{(2)}$INFN, Sezione di Napoli, C. U. Monte S. Angelo, Via Cinthia, $80126$ Napoli, Italy}}}

	\title{A natural cure for causality violations in Newton-Schr\"{o}dinger
		equation}

	\maketitle
	
	
	%\tableofcontents

\begin{abstract}
It is explicitly shown that a one-family parameter model reproducing the
nonlinear Newton-Schr\"{o}dinger equation as the parameter goes to infinity
is free from any causality violation problem for any finite value of it.
This circumstance arises from the intrinsic mechanism of spontaneous state
reduction of the model, absent in the Newton-Schr\"{o}dinger limit. A
specific ideal EPR experiment involving a superposition of two distinct CM's
position states of a massive lump is analyzed, showing recovered
compatibility with QM. Besides, the new framework suggests a soft version of
the Many Worlds Interpretation, in which the typical indiscriminate
proliferation of Everett branches, together with the bizarre inter-branches
communications made possible by nonlinearity, are strongly suppressed in the
macroscopic world by the same mechanism of state reduction.
\end{abstract}

\section{Introduction}

Causality principle lies the foundations of all the Science \cite{pearl}. That a cause precedes its effect is strictly linked, in Special Relativity, to the impossibility of superluminal communications (no-signaling condition). In Quantum Mechanics (QM) this principle is ensured by the kinematic structure of the theory and by the linearity of the dynamics, a circumstance known as `peaceful coexistence between QM and Special Relativity'. As a consequence, entanglement between two space-like separated regions cannot be exploited for superluminal communication. This consideration led some authors to include the no-signaling condition in the basic set of axioms of QM \cite{Simon}. Recently a further generalization has been introduced as well, known as the principle of information causality: it states that there cannot be more information available than was transmitted \cite{paw}.\\
Notwithstanding, the
issue of a possible modification of non-relativistic QM in a nonlinear
sense has a long-standing history, starting from Wigner's suggestion \cite%
{Wigner}. The motivation for this search is twofold. On one side, there was
a need to understand whether linearity is a foundational aspect of QM or
should be considered just an approximation, though a very good one. On the
other side, the measurement problem, and the seemingly subjective border
separating quantum and classical domain, led people to explore nonlinear
modifications of the Schr\"{o}dinger equation. This subject, however, should
not be confused with the issue of the nonlinear versions of the Schr\"{o}dinger equation considered on completely different grounds. In fact there
exists a vast literature on what is called `nonlinear Schr\"{o}dinger
equation', as the famous Gross-Pitaevskii equation describing Bose-Einstein
condensates, which has to be considered as a mean field limit in the
framework of the standard theory and, at the best, only a test-case in
nonlinear dynamics.

Despite some serious efforts to introduce nonlinear corrections \cite%
{WeinbergAnnals} assuming that the observables include all the usual linear
Hermitian operators, stringent limitations on the allowed nonlinearities
became soon evident on the basis of the no-signaling condition
\cite{Gisin, Polchinski}. Nowadays this latter is widely recognized to be a very powerful condition. In fact, for example, it can be used alone to find the
maximum fidelity in the copy of quantum states \cite{BuzekHillery}.

Interestingly, as Polchinski showed \cite{Polchinski}, while one can
construct a theory free from causality problems by means of the stringent
requirement that the Hamiltonian of each subsystem depends (in the formalism
of Ref.\cite{WeinbergAnnals}) only on its density matrix, unusual (Everett)
communications take place among the different branches of the wave function.
That is, the state of a system at any given instant of time and in a
specific world (or mind!) depends also on what happened or \textit{is}
happening in some other worlds (minds). Incidentally, this circumstance
could be used, in principle, to devise an experiment to test the existence
of parallel worlds, independently of a detailed knowledge of human brain
perception's physiology (required, for example, in the conceptual experiment
discussed by Deutsch \cite{Deutsch}).

We note that these foundational problems of QM are not isolated, but appear
in close connection with the still elusive theory of quantum gravity. In
particular, the way in which gravitational fields are produced by quantum
matter is still controversial. Indeed it is not even clear if gravity has to
be quantized at all, in which case the existence of a quantum superposition
of space-times is implied (see \cite{Penrose}
for some related conceptual problems); or gravity is intrinsically
classical, and should be treated consequently. A natural candidate for this
latter hypothesis is the semi-classical gravity, which (at least in its
simplest form) prescribes to take as the source of the field (appearing on
the right hand side of the Einstein field equation) the expectation of the
quantum energy-momentum tensor. Its Newtonian limit, the famous Newton-Schr%
\"{o}dinger equation, turns out to be a non-linear quantum mechanical
equation \cite{Bahrami, Giulini}. While showing some interesting features,
such as the existence of self-localized stationary solutions \cite{Moroz},
this equation is plagued by the causality problems mentioned above \cite%
{Bahrami}.

To give a further chance to semiclassical gravity, various attempts have
been made, inspired and connected with the phenomenological collapse models,
to add \textit{ad hoc} stochastic terms to the Newton-Schr\"{o}dinger term
in such a way that non-classical (delocalized) states become unstable,
rapidly collapsing to well-localized states \cite{Bahrami,Pearle&Squires,Nimm}.

In the present work we take a somehow different viewpoint, by looking at the
Newton-Schr\"{o}dinger equation as a mean field approximation of more
fundamental quantum mechanical equations, technically in a quite similar way
in which Gross-Pitaewski nonlinear equation is derived from standard QM. In
particular, a single-particle \textit{N-S} equation is regarded as the mean
field approximation of an equation of $N$ identical copies of the particle,
gravitationally interacting among them, as $N$ tends to infinity. Indeed, it
was showed that the limit of $N\rightarrow \infty $ gives back the
Newton-Schr\"{o}dinger equation \cite{DeFil+DeFilMaim}. More specifically,
within this model, physical observations are referred to only one of these
particles, while the remaining ones are considered to belong to an \textit{%
hidden} system. The $N$ particles interact uniquely via gravitational
interaction, while the global state of the system is constrained to be
symmetric with respect to particles state permutations. (Incidentally, it is
interesting in this respect the observation by Adler pointing to an
interpretational problem with particles self-interaction within the Hartee
approximation of the Newton-Schr\"{o}dinger equation \cite{Adler}). The evolved
physical state is obtained by tracing out the unobservable degrees of
freedom of the $N-1$ particles (see Appendix \ref{general} for a
self-contained second-quantization general formulation of the model).

The model, known as Nonunitary Newtonian Gravity (NNG, from now on), has
been studied in some detail, in particular the limit $N=2$, showing the
interesting property of (entropic) dynamical self-localization for masses
above the (sharp) threshold of $10^{11}$ proton masses ($m_{p}$ from now on), with
precise signatures susceptible to future experimental tests \cite%
{DeFMaimPRD}, \cite{DeFilMaimRob}, \cite{Nostro}.

Since generally pure states evolve into mixed states even for isolated
systems \cite{DeFMaimPRD},\cite{NostroTwo}, the fundamental description of
physical reality cannot be associated to the wave function. Instead, density
matrix has to be considered as a fundamental description of the Nature (see,
concerning this last point, the latest approach of S. Weinberg to the
foundations of QM \cite{WeinStateVec}).

While it can be inferred directly by the well-posedness of the model that it
is free from causality problems, in the following we will show this fact
explicitly, unfolding its physical basis. In particular we will elucidate
how the basic mechanism operates and guarantees the no-signaling condition.
It turns out that this mechanism is strictly related to the dynamical
mechanism of state reduction naturally embedded within the model and
completely suppressed in the limit $N\rightarrow \infty $ (see Appendix \ref%
{reduction} for an explicit demonstration). This is the same effect
operating in the stochastic versions of Newton-Schr\"{o}dinger, but here it
emerges naturally and is not an \textit{ad hoc} prescription, while state
reduction appears to be a necessary built-in consequence of the no-signaling
condition.

The plan of the present work is as follows. In Section 2, we present in the
simplest possible setting the Newton-Schr\"{o}dinger limit, helping to
elucidate its physical rationale. Then, in Section 3 we analyze a specific EPR situation,
showing how in the simplest $N=2$ case the oddities of causality violations are effectively cured. In Section 4, we show how a communication among
Everett worlds is instead possible, though strongly limited by the model
dynamics itself. Some concluding remarks, in particular on a variant of the
Everett Many Worlds interpretation suggested by our results, end the paper.

\section{Newton-Schr\"{o}dinger limit}

Let's begin by seeing how a superposition of states looks like within the
model. Consider in the ordinary QM setting a superposition of two states of
a body corresponding to two different locations of its CM. Its
representation in the theory goes as follows. Given $p,\,q\,>0$, with $p+q=1$%
, the CM state of the system and of its hidden counterparts, \textit{i.e.}
the \textit{metastate}, is

\begin{equation}
\Vert \Psi ^{(N)}\rangle \rangle =\otimes _{i=1}^{N}(\sqrt{p}|x^{i}\rangle +%
\sqrt{q}|y^{i}\rangle )=\sum_{k=0}^{N}\sqrt{\left(
\begin{array}{c}
N \\
k%
\end{array}%
\right) }p^{\frac{k}{2}}q^{\frac{N-k}{2}}|x\rangle ^{\otimes k}|y\rangle
^{\otimes (N-k)},  \label{metastat}
\end{equation}%
where
\begin{equation*}
|x\rangle ^{\otimes k}|y\rangle ^{\otimes (N-k)}\equiv \frac{1}{\sqrt{N_{P}}}%
\sum_{P}P\Vert x^{1}\dots x^{k}y^{k+1}\dots y^{N}\rangle \rangle .
\end{equation*}%
(A proof of equation (\ref{metastat}) can be found in Appendix A).

Now, starting from the normalization
\begin{equation*}
\langle \langle \Psi ^{(N)}\Vert \Psi ^{(N)}\rangle \rangle
=\sum_{k=0}^{N}\left(
\begin{array}{c}
N \\
k%
\end{array}%
\right) p^{k}q^{N-k},
\end{equation*}%
for $N$ sufficiently large the binomial term can be approximated by a
gaussian
\begin{equation*}
B_{p}(k,N)\sim \aleph (Np,N(1-p)),
\end{equation*}%
and, putting $\alpha =\frac{k}{N}$, we get
\begin{equation}
\langle \langle \Psi ^{(N)}\Vert \Psi ^{(N)}\rangle \rangle =\int_{0}^{1}%
\frac{1}{\sqrt{2\pi p(1-p)/N}}e^{-\frac{(\alpha -p)^{2}}{2p(1-p)/N}}d\alpha %
\xrightarrow[N\rightarrow \infty]{}\int_{0}^{1}\delta (\alpha -p)\,d\alpha .
\end{equation}%
Then, in the limit $N\rightarrow \infty $ the only surviving contribution of
the original sum is that with $k=Np$, \emph{i.e.} the huge superposition of
states reduces to the single (central) term:
\begin{equation*}
\Vert \Psi ^{(N)}\rangle \rangle \xrightarrow[N\rightarrow \infty]{}\Vert
\Psi ^{\infty }\rangle \rangle \propto \lim_{N\rightarrow \infty }|x\rangle
^{\bigotimes Np}|y\rangle ^{\bigotimes N(1-p)}.
\end{equation*}%
In other words, the superposition reduces to the state in which a fraction
\textit{p} of meta-matter is displaced to \textit{x} position and the
remaining part to position \textit{y}.

%(3)---------------------------------------------------------------

\section{Analysis of an ERP situation}

Following the simple argument given in Ref.\cite{Pearle&Squires}, let's
consider a sphere of matter of ordinary density of radius $R$, and let the
state $|Z\rangle $ denote the sphere with center on the z-axis at $z=Z,$ and
suppose the state of the sphere is, within the ordinary QM setting, $(|+Z\rangle
+|-Z\rangle )/\sqrt{2}.$ A probe mass moving along the $x-$ axis will,
according to the non-relativistic quantum theory of gravity, become
entangled with the state of the sphere, resulting in the state vector $%
(|+Z\rangle |up\rangle +|-Z\rangle |down\rangle )/\sqrt{2}$, where $%
|up\rangle $ ($|down\rangle $) means that the probe is deflected in the
positive (negative) $z-$direction. According to semi-classical gravity, the
probe mass should be undeflected. This was experimentally tested under the
hypothesis that QM continues to hold in the macroscopic domain \cite%
{PageGeilker}, with the (not unexpected) result that the mass is deflected.
Indeed, semiclassical gravity implies precisely a modification of QM in this
domain, so the hypothesis is self-contradictory.

Instead, a serious theoretical objection to semiclassical gravity is, as
said above, that it allows superluminal communication and then causality violation. To see this, consider
the entangled state $(|+Z\rangle |0\rangle +|-Z\rangle |1\rangle )/\sqrt{2}$,
where the state $|0\rangle $ and $|1\rangle $ denote orthogonal states of a
two state system (qbit) which is at a large distance from the sphere, but
close to a \textquotedblleft sender\textquotedblright. A probe mass is then
used as before.

If the sender chooses not to measure the system, the ``receiver'', who is
close to the sphere and uses the probe mass as described above, finds it
undeflected. If, on the other hand, the sender chooses to measure the
system, thereby finding it to be in the state $\vert 0\rangle $ or $\vert
1\rangle $, the sphere will immediately be in the state $\vert +Z\rangle $
or $\vert -Z\rangle $ respectively. Then the receiver will be able to see
this because the probe mass will now be deflected up or down.

Let's translate now this conceptual experiment into our model of replicas.

The entangled state of the \textit{sphere+q-bit} is analogous to Eq. (\ref%
{metastat}),
%\begin{widetext}
\begin{eqnarray}\label{metastatq}
\|\Psi^{(N)}\rangle\rangle &=&\bigotimes _{i=1}^N\frac{1}{\sqrt{2}}(\rvert+Z^{i}\rangle\vert 0^{i}\rangle+\rvert-Z^{i}\rangle\vert 1^{i}\rangle) =\notag \\
& = &\sum_{k=0}^{N}\sqrt{\left(\begin{array}{c}N \\k\end{array}\right)}\biggl(\frac{1}{2}\biggr)^{N/2}|+Z;0\rangle^{\otimes k} |-Z;1\rangle^{\otimes (N-k)},
\end{eqnarray}
%\end{widetext}
where the tensor product terms are defined in the same way as in Eq. (\ref%
{metastat}).

Consider now the probe particle (supposed to have a mass much more smaller
than the lump) shut just over the superposition. Including the probe in the
system's description, we have that an initially unentangled (global)
state of the probe and of the composite \textit{sphere+q-bit} system evolves
towards an entangled one,
%\begin{widetext}
\begin{eqnarray}\label{product}
\Vert \varphi_I^{(N)}\rangle\rangle\bigotimes \Vert \Psi^{(N)}\rangle\rangle \xrightarrow[time\mbox{ }evolution]{} \sum_{k=0}^{N} \sqrt{\left(\begin{array}{c}N \\k\end{array}\right)}\biggl(\frac{1}{2}\biggr)^{N/2}\Vert \varphi_k^{(N)}\rangle\rangle\vert +Z;0\rangle^{\bigotimes k}\vert -Z;1\rangle^{\bigotimes (N-k)},
\end{eqnarray}
%\end{widetext}
where $\Vert \varphi _{I}^{(N)}\rangle \rangle =\bigotimes_{m=1}^{N}|\varphi
_{I}^{m}\rangle $ is the product of identical copies of the probe's state
(it is assumed that probe's mass is so small that gravity-induced internal
entanglement among copies is irrelevant; see the example of the particle in
Earth's gravitational field in Ref.\cite{DeFMaimPRD}). The meaning of the
index $k$ is clarified in Fig. 1.
%\begin{widetext}
\begin{figure}[htbp]
\centering
  \includegraphics[scale=0.5]{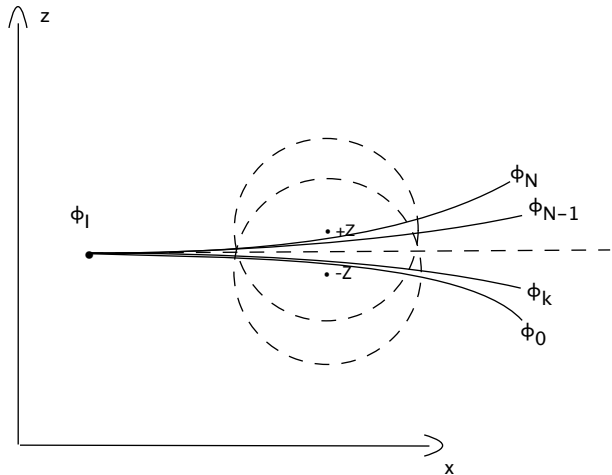}\\
\caption{\small{Seen from above, deflection branches of the probe particle passing over the massive sphere in superposition of Eq. (\ref{product}).}} \label{fig:deflection}
\end{figure}
%\end{widetext}

For simplicity let's consider the simplest case with $N=2$, for which we
have only three branches in the superposition (downward, central and upward
trajectories). Generalization to a generic $N$ is straightforward.

First of all, observe that in order to detect a whatever deflection in the
probe particle's trajectory, the size of the wave packets describing the
particle's states in the superposition should be smaller than the deflection
itself. Moreover, their spreading along the path have to be taken into
account before the position measurement of the particle's position along the
\textit{z}-axis. Denoting with $v_{x}$ the velocity of the particle, $m$ its
mass, and $M$ the mass of the sphere, the time during which the sphere's
gravitational attraction is effective is of the order of $T\sim R/v_{x}.$
Then it should be
\begin{equation*}
\frac{GMT}{R^{2}}\gtrsim \dfrac{\hbar }{mZ}.
\end{equation*}
\begin{figure}[tbph]
	\centering
	\includegraphics[scale=1]{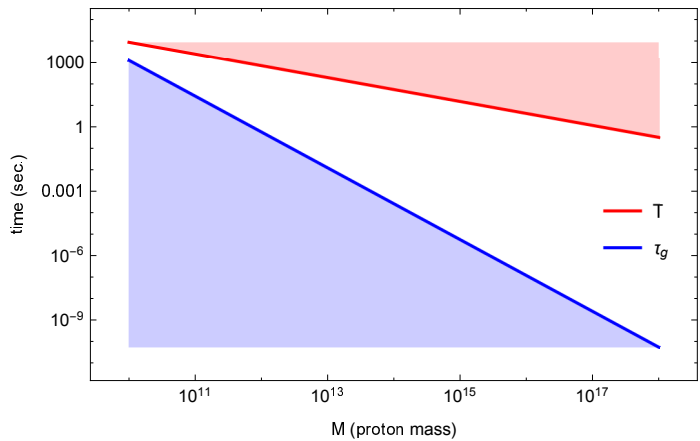}
	\caption{{\protect\small {White area denotes the `interdiction' zone in the
				diagram (\textit{M, time}), for which superluminal communications would be
				possible.  $T$ and $\tau_g$, representing the upper and lower lines respectively,
				are defined within the main text.}}}
	\label{fig:interdition}
\end{figure}
On the other hand, a peculiar dynamical signature of the models (studied in
Ref.\cite{DeFMaimPRD}; see also the calculation below in the limiting case $%
N=2$) is that, for a lump of mass \textit{M} above a threshold of $10^{11} m_{p}$, a superposition of the CM wave packets separated by a
distance much smaller than the body's size undergoes a rapid state reduction
after a characteristic time $\tau _{g}\sim \hbar G^{-1}M^{-5/3}\rho ^{-1/3}$
(see Appendix \ref{reduction} for a self-contained derivation), leading from
the initial superposition to an ensemble of localized states (see footnote~\ref{nota} on page~\pageref{nota}). The additional condition for the detection of a significant deflection of
the probe is that the measurement time should be smaller than this reduction
time, which together with the constraint above gives the set of conditions
that must be satisfied simultaneously:

\begin{equation}
\left\{
\begin{array}{ll}
T & \gtrsim \biggl(\dfrac{\hbar }{m}\biggr)^{1/3}\biggl[\frac{1}{2}\biggl(%
\dfrac{4}{3}\pi \rho \biggr)^{\frac{2}{3}}GM^{\frac{1}{3}}\biggr]^{-\frac{2}{%
3}} \\
T & \ll \tau _{g}=\hbar G^{-1}M^{-5/3}\rho ^{-1/3}.
\end{array}%
\right.   \label{conditions}
\end{equation}%
As said previously, the generalization to a generic \textit{N} is
straightforward. In other words, our conclusion can be stated by saying that
gravity-induced state reduction is so rapid to forbid a sufficiently long
measurement, which would otherwise permit a deflection discrimination with
respect to the spreading of the wave packet. These conditions have been
depicted in Fig. \ref{fig:interdition}, from which it is clear that
they cannot be satisfied together.

In other words the `peaceful coexistence' between (deterministic) QM and
special relativity indeed \textit{imply} linearity (see for example \cite%
{Simon}), unless a certain amount of predictability loss is present.

%(4)---------------------------------------------------------------

\section{Everett phone}

As mentioned above, one of the drawbacks of the introduction of
nonlinearities in a theory with a properly restricted observational algebra
is the appearance of bizarre communications among Everett branches of the
wave function. This circumstance appears also within the nonlinear model
described here. At variance with the other nonlinear modifications of the
theory, in this case predictability loss strongly limits such communication
possibilities as far as macroscopic (mass) cat states are involved, as we
are going to show. Let's start by looking at the interaction among branches;
then we turn to the possibility to use these interactions to construct, in
principle, an Everett phone.

The existence of effective interactions among Everett branches can be
immediately inferred by considering the following physical situation.
Assume an initial superposition of localized states of an isolated lump of
matter (whose mass we will assume just above the localization mass
threshold). Branches or worlds independence means that, following the
evolution of each of the states independently from each other and, then,
forming the final superposition of the evolved states at some later time, is
equivalent to considering the time evolution of the global state from the
beginning. This is clearly not true within the present model, since the
state reduction dynamics rely precisely on the existence of all the other
states in the superposition, while the exact reduction time depends also on
the spatial distribution of the localized states. (For example, a long
cigar-shaped matter distribution would be reduced more slowly in comparison
with a ball-like distribution occupying the same volume.)

Said in other words, the evolution of a superposition of localized states
does not coincide with the superposition of the evolved localized states.
This is a statement of the existence of interactions among the branches.
Incidentally, we stress that Everett branches should not be confused with
the copy/copies of the physical system, which represent just a useful and
easy way to formulate the model.

Let's now illustrate, following Polchinski's proof of principle, how an
Everett phone could be constructed on the basis\ of the theory. Consider a
two-level system $S$ initially in the state $|\Phi _{0}\rangle $ and a
system $M$, initially in the state $|M_{0}\rangle $, that we can imagine
composed in general of \textit{environment} + \textit{recording device} +
(eventually) \textit{an observer's mind}.\newline
Measuring the q-bit state in the $\{|1\rangle ,|0\rangle \}$ basis, we
get
\begin{equation}
|\Phi _{0}\rangle \otimes |M_{0}\rangle \longrightarrow \frac{1}{\sqrt{2}}%
\biggl(|1\rangle |M_{1}\rangle +|0\rangle |M_{2}\rangle \biggr).
\label{statoin}
\end{equation}%
Calling $\sigma _{3}$ the third Pauli matrix, if we obtain $1$, it is $\rho
^{+}=\frac{1+\sigma _{3}}{4}$; otherwise $\rho ^{-}=\frac{1-\sigma _{3}}{4}$.
In this second case, the observer (or the recording automatic system) can
follow one of two actions: (\textit{a}) nothing; (\textit{b}) rotates the
q-bit state into the $\widehat{\mathbf{i}}$ direction. We analyze these cases
in the NNG representation, starting from the second case.

\begin{itemize}
\item \textit{Case} (\textit{b}).
\end{itemize}

We write our (meta-)state $||\psi(t)\rangle\rangle^{(I)}$ as
\begin{eqnarray}
&&\frac{1}{2}\biggl(|1M_{1}\rangle +|\mathbf{i}M_{2}\rangle \biggr)\otimes %
\biggl(|\widetilde{1M_{1}}\rangle +|\widetilde{\mathbf{i}M_{2}}\rangle %
\biggr)  \notag  \label{metastatobt} \\
&=&\frac{1}{2}\biggl(|1\rangle |\widetilde{1}\rangle |M_{1}\rangle |%
\widetilde{M_{1}}\rangle +|1M_{1}\rangle |\widetilde{\mathbf{i}M_{2}}\rangle
+|\mathbf{i}\rangle |\widetilde{1}\rangle |M_{2}\rangle |\widetilde{M_{1}}%
\rangle +|\mathbf{i}\rangle |\widetilde{\mathbf{i}}\rangle |M_{2}\rangle |%
\widetilde{M_{2}}\rangle \biggr)\notag \\
&\xrightarrow[time\mbox{ }evolution]{} &\frac{1}{2}\biggl(|1M_{1};\widetilde{1M_{1}}\rangle
_{t}+|1M_{1};\widetilde{\mathbf{i}M_{2}}\rangle _{t}\biggr).
\end{eqnarray}%
(Remember that $|\mathbf{i}\rangle =\frac{1}{\sqrt{2}}\bigl(|1\rangle
+i|0\rangle \bigr)$.) Suppose that, at the beginning, state $|1\rangle $ was
measured; in (\ref{metastatobt}), that part of\ the wave function not living
in the branch of $M_{1}$ has to be disregarded. Writing the most general
time-evolved relevant states in the form
\begin{equation*}
|1M_{1};\widetilde{1M_{1}}\rangle _{t}=\sum_{(i,\mu )(\tilde{\imath},\tilde{%
\mu})}c_{(i,\mu )(\tilde{\imath},\tilde{\mu})}^{(1)}|i\mu ;\tilde{\imath}%
\tilde{\mu}\rangle _{0}
\end{equation*}%
and
\begin{equation*}
|1M_{1};\widetilde{\widehat{\mathbf{i}}M_{2}}\rangle _{t}=\sum_{(i,\mu )(%
\tilde{\imath},\tilde{\mu})}c_{(i,\mu )(\tilde{\imath},\tilde{\mu}%
)}^{(2)}|i\mu ;\tilde{\imath}\tilde{\mu}\rangle _{0}\ ,
\end{equation*}%
the general expression for the evolved q-bit state is ($\widetilde{S}$ is the
hidden q-bit)

\begin{eqnarray}
\rho (t) &=&\overset{}{\underset{M,\ \widetilde{S}}{Tr}}\biggl[\Vert \psi
(t)\rangle \rangle ^{(I)}\langle \langle \psi (t)\Vert \biggr]=\sum_{\alpha
,(\tilde{j},\tilde{\alpha})}\langle \alpha ,\tilde{j},\tilde{\alpha}\Vert
\psi (t)\rangle \rangle ^{(I)}{}\langle \langle \psi (t)\Vert \alpha ,\tilde{%
j},\tilde{\alpha}\rangle =  \notag  \label{traccia} \\
&=&\frac{1}{4}\sum_{i,i^{\prime }}\biggl[\sum_{\alpha ,(\tilde{j},\tilde{%
\alpha})}\biggl(c_{(i,\alpha )(\tilde{j},\tilde{\alpha})}^{(1)}c_{(i^{\prime
},\alpha )(\tilde{j},\tilde{\alpha})}^{(1)\ast }+c_{(i,\alpha )(\tilde{j},%
\tilde{\alpha})}^{(2)}c_{(i^{\prime },\alpha )(\tilde{j},\tilde{\alpha}%
)}^{(2)\ast }+c_{(i,\alpha )(\tilde{j},\tilde{\alpha})}^{(1)}c_{(i^{\prime
},\alpha )(\tilde{j},\tilde{\alpha})}^{(2)\ast }+c_{(i,\alpha )(\tilde{j},%
\tilde{\alpha})}^{(2)}c_{(i^{\prime },\alpha )(\tilde{j},\tilde{\alpha}%
)}^{(1)\ast }\biggr)\biggr]|i\rangle \langle i^{\prime }|=  \notag \\
&\equiv &\dfrac{1}{4}\sum_{i,i^{\prime }}C_{i,i^{\prime }}|i\rangle \langle
i^{\prime }|\ ,
\end{eqnarray}

from which

\begin{equation}
\langle \sigma _{3}\rangle =Tr\biggl[\sigma _{3}\rho _{s}^{(t)}\biggr]%
=C_{1,1}-C_{2,2}.  \label{spinmedio}
\end{equation}

\begin{itemize}
\item \textit{Case} (\textit{a}).
\end{itemize}

In this case our meta-state is
\begin{equation*}
\frac{1}{2}\biggl(|1M_{1}\rangle +|0M_{2}\rangle \biggr)\otimes \biggl(|%
\widetilde{1M_{1}}\rangle +|\widetilde{0M_{2}}\rangle \biggr).
\end{equation*}%
With a bit of algebra we can verify that the above expression remains\
formally unchanged, but with values of $c_{(i,\alpha )(\tilde{j},\tilde{%
\alpha})}^{(2)}$ generally changed.

To summarize the above scheme, the observer measuring $0$ at the beginning is
able to send a bit of information (say, `$a$' if he chooses action (\textit{a%
}), `$b$' if he chooses action (\textit{b})) to the observer who originally
measured $1$. In this way a procedure to communicate between two Everett
branches is set up.

In order to ensure the consistency of our argument, we can verify that in
the absence of gravitational interactions, case (\textit{a}) and case (%
\textit{b}) coincide. For this purpose it is convenient to use `mixed basis
states' (in which one factor of the product tensor is taken at the initial
time as before, while the other is considered at time $t$) to represent the
time evolved states, \textit{i.e.} writing

\begin{equation*}
|1\ M_{1};\widetilde{1\ M_{1}}\rangle _{t}=\sum_{(i,\mu )(\tilde{\imath},%
\tilde{\mu})}c_{(i,\mu )(\tilde{\imath},\tilde{\mu})}^{(1)}|i\mu \rangle _{0}%
\underset{=e^{-\frac{i\tilde{H_{0}}t}{\hbar }}|\tilde{\imath},\tilde{\mu}%
\rangle _{0}}{\underbrace{|\tilde{\imath},\tilde{\mu}\rangle _{t}}}
\end{equation*}%
and
\begin{equation*}
|1\ M_{1};\widetilde{\mathbf{i}\ M_{2}}\rangle _{t}=\sum_{(i,\mu )(\tilde{%
\imath},\tilde{\mu})}c_{(i,\mu )(\tilde{\imath},\tilde{\mu})}^{(2)}|i\mu
\rangle _{0}|\tilde{\imath},\tilde{\mu}\rangle _{t}\ .
\end{equation*}
Remembering that ${}_{0}\langle \widetilde{i\mu }|\widetilde{1M_{2}}\rangle
_{0}={}_{t}\langle \widetilde{i\mu }|\widetilde{1M_{2}}\rangle _{t}=\delta _{%
\widetilde{i\mu }}\delta _{\widetilde{1M_{2}}}$, it is immediate to prove
the identity of $\langle \sigma_{3}\rangle $ for the two cases, given by

\begin{equation*}
\langle \sigma _{3}\rangle =2\sum_{\alpha }\biggl[\biggl\vert{}_{0}\langle
1,\alpha |1,M_{1}\rangle _{t}\biggr\vert^{2}-\biggl\vert{}_{0}\langle
2,\alpha |1,M_{1}\rangle _{t}\biggr\vert^{2}\biggr].
\end{equation*}

It's important to note that, apart from formal consistency, the above result
show that an Everett phone cannot work for truly microscopic systems.

As the last point concerning branch communications, it's easy to see that if
the superposition $M_{1}+M_{2}$ involves sufficiently spatially-separated
massive systems, then the expectation values $\langle \sigma _{3}\rangle ~$%
for case (\textit{a}) and (\textit{b}) become identical in a very short
time, meaning that after that time Everett Universes' communication
possibilities are strongly suppressed.

Suppose, in fact, that $M_{1}$ and $M_{2}$ are approximate CM's position
eigenstates with a mass above threshold $m\gtrsim 10^{12} m_{p}$ at
a distance $\sim \left( m_{p}/m\right) ^{1/2}cm$ from each other\footnote{
By `superposition of
two approximate position eigenstates' we have meant,
throughout the text, a superposition of two contiguous clusters of
localized states; as a matter of fact, a necessary condition for a complete
state reduction within the characteristic time $\tau _{g}$ is that the
superposition is composed by a large number of localized states, of width $%
\sim \left( m_{p}/m\right) ^{1/2}cm$ (see \cite{DeFilMaimRob} for an
explicit numerical simulation of this case). Otherwise, a superposition of
two really separated localized states would lead to a rapid oscillation of
coherences in the basis of positions. Indeed, for all practical purposes,
for sufficiently massive bodies, Nonunitary Gravity acts to reduce the
overall quantum state at the very beginning of the (unitary) process of
superposition formation.\label{nota}}. As before, the dynamics in this situation lead to
a rapid reduction of the state within the characteristic time $\tau _{g}$.
For the sake of simplicity, let's suppose that the spin dynamics is
slower than $\tau _{g}$; then, with respect to the case (\textit{b}), the
state $\frac{1}{\sqrt{2}}(|1M_{1}\rangle +|\widehat{\mathbf{i}}M_{2}\rangle )
$ evolves in a time $t\gtrsim \tau _{g}$ to give the almost diagonal state
\begin{equation*}
\rho (t)\simeq \frac{1}{2}\biggl(|1M_{1}\rangle \langle 1M_{1}|+|\widehat{%
\mathbf{i}}M_{2}\rangle \langle \widehat{\mathbf{i}}M_{2}|\biggr)\ ,
\end{equation*}%
from which, after tracing out system $M$, we see that there is a given
probability to obtain $1$ associated to $M_{1}$, and the complementary
probability to obtain $\widehat{\mathbf{i}}$ associated to (mind, recording
device, etc.) $M_{2}$ without any `transfer' from branch of $M_{2}$ to
branch $M_{1}$ after a time $t\gtrsim \tau _{g}$.

In conclusion, to set-up an efficient Everett communication, we need
macroscopic mass superpositions (because gravity is responsible for
nonlinearity, the basic ingredient for Everett communication!). At the same
time, the macroscopic nature of the required quantum superposition implies
its rapid decay, putting a severe trade-off between efficiency and practical
feasibility of the Everett phone.

\section{Conclusions}

We have explicitly shown how the well-known causality problems of the Newton-Schrödinger equation can be cured in a natural and satisfactory way at the price of introducing a certain amount of
predictability loss into the theory. The character of this extra-level of
indeterminism is such that it is completely irrelevant for microscopic
systems, while prompting large entropy production when macroscopic bodies
are involved. Being the ensuing theory, equivalent to NNG, a fully
consistent QM model, we have shown in an explicit physical setting how the
theory itself gets rid automatically of the superluminal communication
channels. This has been accomplished by analyzing an (ideal) EPR-type
experiment involving the superposition of two distinct CM position states of
a massive body. It has been shown that this circumstance arises from the
intrinsic mechanism of spontaneous state reduction of the model, which is
completely suppressed in the Newton-Schr\"{o}dinger limit. As a matter of
fact, the present analysis can be considered as an independent argument for
the introduction of NNG.

As to the experimental consequences of our calculations, one could not design an experimental setting to detect the average gravitational field of a whatever quantum superposition of localized states, even if semiclassical gravity is the correct description for all practical purposes. Turning it the other way around, one could not use semiclassical gravity to design an apparatus to send faster-than-light signals.

Besides, since within NNG density matrix plays a fundamental role, amounting
to the most complete characterization of a physical system's state, the
Everett Many World Interpretation appears to be the most natural conceptual
framework of that theory.

As in other approaches in which non-linearity was introduced at a
fundamental level in QM, the possibility of constructing an Everett phone
connecting different branches of the wave function emerges, though the
mechanism it is based on appears to be strongly inhibited. In fact the
theory gets rid of the huge number of branches continuously forming by
turning each time the macroscopic states superposition into ensembles of
localized states through gravitational self-interaction.

The severe restriction to branching proliferation would be responsible for
keeping the process confined within the microscopic-to-mesoscopic realm. To
further clarify this point, let's consider the famous Schr\"{o}dinger's cat
thought experiment in whatever of its countless versions. The main point
concerning this thought experiment is that a microscopic system's dynamics
(described by the Schr\"{o}dinger equation) is amplified up to the level of
a macroscopic object (like a gun or a bottle of poison), and from it to a
poor cat, turning the world split into two macroscopically well distinct
branches: one in which the cat stands up happy and alive, and one in which
the same cat is lying stretched out on the floor. Decoherence theory adds to
this picture the participation of the surrounding environment in this
branching, but the principle remains exactly the same. What happens when
Nonunitary Gravity is acting? That as soon as the gun's trigger (or the
poison vapor) begins to form superpositions of mesoscopically different mass
displacement states, the overall (cat+killing apparatus+microscopic
system+surrounding environment) pure state is rapidly converted into an
ensemble of classical-like states, meaning that each state has its own
probability of occurrence. It is important to stress that this would not
correspond to the subjective coarse graining, usually given by tracing out
the environmental degrees of freedom. It would, instead, be associated with
a fundamental coarse graining, the adjective `fundamental' meaning that one
cannot resort even in principle to a more complete description, regardless
of how precise and technologically sophisticated his measuring apparatuses
would be.

Notwithstanding, of course in the microscopic realm the branching of the
wave function continues undisturbed, while some remote echoes of other
parallel worlds could eventually take a role in the biological processes
involved in the brain's functioning.

As a final remark of moral order, a \textit{soft} version of the Everett
formulation like the one just described, if found to conform to physical
reality, would presumably sound less horrible, and even
psychologically/morally more acceptable than the quite disturbing notion of
an infinite number of replicas of ourselves doing who knows what, who knows
where.

\textit{Note added}. During the completion of this paper we became aware of
a work by S. De Filippo, which also points out the ability of NNG model in avoiding
superluminal communications \cite{NotaDeFilippo}.

%---------------------------------------------------------------
\begin{appendices}
%\renewcommand{\thesection}{\arabic{chapter}.\arabic{section}}
%\appendix
\numberwithin{equation}{section}
%---------------------------------------------------------------

\section{Proof of Eq. (\protect\ref{metastat})}

Let's proceed by induction. For $N=1,$ Eq. \ref{metastat}) is trivially
verified. Assuming that it is verified for $N,$ it is sufficient to prove
that it is verified for $N+1$, \textit{i.e.} we have to prove the
equivalence of
\begin{equation}
	\biggl(\sum_{k=0}^{N}\sqrt{\binom{N}{k}}p^{\frac{k}{2}}q^{\frac{N-k}{2}%
	}\vert x\rangle ^{\otimes k} \vert y\rangle ^{\otimes (N-k)}\biggr)\times (%
	\sqrt{p}\vert x^{N+1}\rangle +\sqrt{q}\vert y^{N+1}\rangle)
\end{equation}
with
\begin{equation}
	\sum_{k=0}^{N+1}\sqrt{\binom{N+1}{k}}p^{\frac{k}{2}}q^{\frac{N-k+1}{2}}\frac{
		1}{\sqrt{N p^{\prime \prime }}}\sum_{P^{\prime \prime }}P^{\prime \prime
	}\Vert x^1,\dots ,x^{k}, y^{k+1},\dots ,y^{N+1}\rangle\rangle .
\end{equation}
Both expressions are decomposable in a unique way as sums of terms each with
a fixed number $k$ of $x$. So we can show the equivalence of the two
expressions term by term:
\begin{eqnarray*}
	&&\sqrt{\binom{N}{k}}p^{\frac{k}{2}}q^{\frac{N-k+1}{2}}\frac{1}{\sqrt{N p}}
	\sum_{P}P\Vert x^{1},\dots ,x^{k}, y^{k+1},\dots ,y^{N}\rangle\rangle \vert
	y^{N+1}\rangle \\
	& + & \sqrt{\binom{N}{k-1}}p^{\frac{k}{2}}q^{\frac{N-k+1}{2}}\frac{1}{\sqrt{
			N p^{\prime }}}\sum_{P^{\prime }}P^{\prime} \Vert x^1,\dots ,x^{k-1},
	y^{k},\dots ,y^{N}\rangle\rangle \vert x^{N+1}\rangle \\
	& = & \sqrt{N p^{\prime \prime }} p^{\frac{k}{2}}q^{\frac{N-k+1}{2}}\frac{1}{
		\sqrt{N p^{\prime \prime }}}\sum_{P^{\prime \prime }}P^{\prime \prime
		1},\dots ,x^{k}, y^{k+1},\dots ,y^{N+1}\rangle\rangle \vert y^{N+1}\rangle \\
	& \equiv & \sqrt{\binom{N+1}{k}}p^{\frac{k}{2}}q^{\frac{N-k+1}{2}}\vert
	x\rangle^{\otimes k} \vert y\rangle^{\otimes (N+1-k)}.
\end{eqnarray*}
The total number of permutations involved in the above transformations is
equal to
\begin{equation*}
	N_{p}+N_{p^{\prime }}\equiv \binom{N}{k}+\binom{N}{k-1}=\binom{N+1}{k}\equiv
	N p^{\prime \prime }.
\end{equation*}

%---------------------------------------------------------------

\section{General formulation of the model}

\label{general} %\setcounter{equation}{0}
For a more general formulation of the model, it is convenient to switch to
second quantization. Let $H[\psi ^{\dagger },\psi ]$ denote the second
quantized non relativistic Hamiltonian of a finite number of particle
species, like electrons, nuclei, ions, atoms and/or molecules, according to
the energy scale. For notational simplicity, $\psi ^{\dagger },\,\psi $
denote the whole set $\psi _{j}^{\dagger }(x),\,\psi _{j}(x)$ of
creation-annihilation operators, \textit{i.e.} one couple per particle
species and spin component. This Hamiltonian includes the usual
electromagnetic interaction accounted for in atomic and molecular physics.
To incorporate gravitational interactions including self-interactions, we
introduce a color quantum number $\alpha =1,2,\dots ,N$, in such a way that
each couple $\psi _{j}^{\dagger }(x),\,\psi _{j}(x)$ is replaced by $N$
couples $\psi _{j,\alpha }^{\dagger }(x),\,\psi _{j,\alpha }(x)$ of
creation-annihilation operators. The overall Hamiltonian, including
gravitational interactions and acting on the tensor product $\otimes
_{\alpha }F_{\alpha }$ of the Fock space of the $\psi _{\alpha }$ operators,
is then given by
\begin{equation}
	H_{G}=\sum_{\alpha =1}^{N}H[\psi ^{\dagger },\psi ]-\frac{G}{N-1}%
	\sum_{j,k}m_{j}m_{k}\sum_{\alpha <\beta }\int dx\,dy\frac{\psi _{j,\alpha
		}^{\dagger }(x),\,\psi _{j,\alpha }(x)\psi _{k,\beta }^{\dagger }(y),\,\psi
		_{k,\beta }(y)}{|x-y|},
\end{equation}%
where here and henceforth Greek indices denote color indices, $\psi _{\alpha
}\equiv (\psi _{1,\alpha },\psi _{2,\alpha },\dots ,\psi _{N,\alpha })$ and $%
m_{i}$ denotes the mass of the $i-$th particle species, while $G$ is the
gravitational constant. While the $\psi _{\alpha }$ operators obey the same
statistics as the original operators $\psi $, we take advantage of the
arbitrariness pertaining to distinct operators and, for simplicity, we chose
them commuting with one another: $\alpha \neq \beta \Rightarrow \lbrack \psi
_{\alpha },\psi _{\beta }]_{-}=[\psi _{\alpha },\psi _{\beta }^{\dagger
}]_{-}=0.$ The metaparticle state space $S$ is identified with the subspace
of $\otimes _{\alpha }F_{\alpha }$ including the metastate obtained from the
vacuum $\Vert 0\rangle \rangle =\otimes _{\alpha }|0\rangle _{\alpha }$
applying operators built in terms of the product $\prod_{\alpha =1}^{N}\psi
_{j,\alpha }^{\dagger }(x_{\alpha })$ and symmetrical with respect to
arbitrary permutations of the color indices, which, as a consequence, for
each particle species, have the same number of metaparticles of each color.
This is a consistent definition since the time evolution generated by the
overall Hamiltonian is a group of (unitary) endomorphism of $S$. If we
prepare a pure $n-$particle state, represented in the original setting,
excluding gravitational interactions, by
\begin{equation*}
	|g\rangle \doteq \int d^{n}x\,g(x_{1},x_{2},\dots ,x_{n})\psi
	_{j_{1}}^{\dagger }(x_{1}),\psi _{j_{2}}^{\dagger }(x_{2})\dots ,\psi
	_{j_{n}}^{\dagger }(x_{n})|0\rangle ,
\end{equation*}%
its representative in $S$ is given by the metastate
\begin{equation*}
	\Vert g^{\otimes N}\rangle \rangle \doteq \prod_{\alpha }\biggl[\int
	d^{n}x\,g(x_{1},x_{2},\dots ,x_{n})\psi _{j_{1},\alpha }^{\dagger
	}(x_{1}),\psi _{j_{2},\alpha }^{\dagger }(x_{2})\dots ,\psi _{j_{n},\alpha
	}^{\dagger }(x_{n})\biggr]\Vert 0\rangle \rangle .
\end{equation*}

As for the physical algebra, it is identified with the operator algebra of
say the $\alpha =1$ metaworld. In view of this, expectation values can be
evaluated by previously tracing out the unobservable operators, namely with $%
\alpha >1$, and then taking the average of an operator belonging to the
physical algebra. It should be made clear that we are not prescribing an
\textit{ad hoc} restriction of the observable algebra. Once the constraint
restricting $\otimes _{\alpha }F_{\alpha }$ to $S$ is taken into account,
in order to get an effective gravitational interaction between particles of
one and the same color, the resulting state space does not contain states
that can distinguish between operators of different color. The only way to
accommodate a faithful representation of the physical algebra within the
metastate space is to restrict the algebra to that of $\psi _{1}$ operators.
Note that the resulting constrained\ theory is, by construction, a fully
consistent QM theory.
%---------------------------------------------------------------

\subsection{State reduction}

\label{reduction} %\setcounter{equation}{0}
The evolution operator in the interaction representation mapping an initial
physical state $\rho (0)=|\Phi (0)\rangle _{1}\left\langle \Phi
(0)\right\vert $ into the evolved physical state $\rho (t)$ can be written,
according to \cite{DeFil+DeFilMaim, NotaDeFilippo}, as
\begin{eqnarray}
	M(t) &=&\int \mathcal{D}[\varphi ]\prod_{\alpha }\mathcal{D}[\varphi
	_{\alpha }]\mathcal{D}[\varphi ^{\prime }]\prod_{\alpha }D[\varphi ^{\prime }%
	{_{\alpha }}]exp{\biggl[\frac{ic^{2}}{2\hbar }\int dt\,dx\bigl[\varphi
		\nabla ^{2}\varphi -\sum_{\alpha }\varphi _{\alpha }\nabla ^{2}\varphi
		_{\alpha }-\varphi ^{\prime 2}\varphi ^{\prime }+\sum_{\alpha }\varphi
		_{\alpha }^{\prime }\nabla ^{2}\varphi _{\alpha }^{\prime }\bigr]}\biggr]
	\notag  \label{MdiT} \\
	&\times &\biggl[\bigotimes_{\alpha =2}^{N}\;{}_{\alpha }\langle \Phi (0)|%
	\biggr]T^{-1}\exp {\biggl[i\frac{2mc}{\hbar }\sqrt{\frac{\pi G}{N-1}}%
		\sum_{\alpha =2}^{N}\int dt\,dx[\varphi ^{\prime }(x,t)+\varphi _{\alpha
		}^{\prime }(x,t)]\psi _{\alpha }^{\dagger }(x,t)\psi _{\alpha }(x,t)\biggr]}
	\notag \\
	&\times &T\exp {\biggl[-i\frac{2mc}{\hbar }\sqrt{\frac{\pi G}{N-1}}%
		\sum_{\alpha =2}^{N}\int dt\,dx[\varphi (x,t)+\varphi _{\alpha }(x,t)]\psi
		_{\alpha }^{\dagger }(x,t)\psi _{\alpha }(x,t)\biggr]}\bigotimes_{\alpha
		=2}^{N}|\Phi (0)\rangle _{\alpha } \\
	&\times &T\exp {\biggl[-i\frac{2mc}{\hbar }\sqrt{\frac{\pi G}{N-1}}\int
		dt\,dx[\varphi (x,t)+\varphi _{1}(x,t)]\psi _{1}^{\dagger }(x,t)\psi
		_{1}(x,t)\biggr]}|\Phi (0)\rangle _{1}  \notag \\
	&\times &{}_{1}\langle \Phi (0)|T^{-1}\exp {\biggl[i\frac{2mc}{\hbar }\sqrt{%
			\frac{\pi G}{N-1}}\int dt\,dx[\varphi ^{\prime }(x,t)+\varphi _{1}^{\prime
		}(x,t)]\psi _{1}^{\dagger }(x,t)\psi _{1}(x,t)\biggr]}.  \notag
\end{eqnarray}%
Let's consider localized states $|z\rangle _{1}$, which are approximate
eigenstates of the density operator $\psi _{1}^{\dagger }(x,t)\psi
_{1}(x,t)|z\rangle _{1}\approx n(x-z)|z\rangle _{1}$ that are quasi
stationary, apart from a slow spreading proportional to $N$ (associated to
the center of metamass $Nm$ spreading). Taking as initial state a
superposition of a large number $\mathcal{N}$ of localized states,
\begin{equation*}
	|\Phi (0)\rangle _{1}=\frac{1}{\sqrt{\mathcal{N}}}\sum_{j=1}^{\mathcal{N}%
	}|z_{j}\rangle _{1},
\end{equation*}%
where $\mathcal{N}$ is the number of states in superposition. %from the
%result of Appendix A we have that
%\begin{eqnarray}  \label{lunga}
%{}_1\langle z_h\vert M(t)\vert z_k\rangle _1 = \frac{1}{\mathcal{N}}e^{i/%
%\mathcal{N}\sum_{j=1}^{\mathcal{N}}}\mathcal{A}_j.
%\end{eqnarray}

We want to evaluate explicitly the matrix element
\begin{eqnarray*}
	&&{}_{1}\langle z_{h}|M(t)|z_{k}\rangle _{1} \\
	&=&\int \mathcal{D}[\varphi ]\prod_{\alpha }\mathcal{D}[\varphi _{\alpha }]%
	\mathcal{D}[\varphi ^{\prime }]\prod_{\alpha }D[\varphi ^{\prime }{_{\alpha }%
	}]exp{\biggl[\frac{ic^{2}}{2\hbar }\int dt\,dx\bigl[\varphi \nabla
		^{2}\varphi -\sum_{\alpha }\varphi _{\alpha }\nabla ^{2}\varphi _{\alpha
		}-\varphi ^{\prime 2}\varphi ^{\prime }+\sum_{\alpha }\varphi _{\alpha
		}^{\prime }\nabla ^{2}\varphi _{\alpha }^{\prime }\bigr]}\biggr] \\
	&\times &\biggl[\bigotimes_{\alpha =2}^{N}\frac{1}{\sqrt{\mathcal{N}}}%
	\sum_{j=1}^{\mathcal{N}}\,{}_{\alpha }\langle z_{j}|\biggr]T^{-1}\exp {%
		\biggl[i\frac{2mc}{\hbar }\sqrt{\frac{\pi G}{N-1}}\sum_{\alpha =2}^{N}\int
		dt\,dx[\varphi ^{\prime }(x,t)+\varphi _{\alpha }^{\prime }(x,t)]\psi
		_{\alpha }^{\dagger }\psi _{\alpha }(x,t)\biggr]} \\
	&\times &T\exp {\biggl[-i\frac{2mc}{\hbar }\sqrt{\frac{\pi G}{N-1}}%
		\sum_{\alpha =2}^{N}\int dt\,dx[\varphi (x,t)+\varphi _{\alpha }(x,t)]\psi
		_{\alpha }^{\dagger }(x,t)\psi _{\alpha }(x,t)\biggr]}\biggl[%
	\bigotimes_{\alpha =2}^{N}\frac{1}{\sqrt{\mathcal{N}}}\sum_{j=1}^{\mathcal{N}%
	}|z_{j}\rangle _{\alpha }\biggr] \\
	&\times &{}_{1}\langle z_{h}|T\exp {\biggl[-i\frac{2mc}{\hbar }\sqrt{\frac{%
				\pi G}{N-1}}\int dt\,dx[\varphi (x,t)+\varphi _{1}(x,t)]\psi _{1}^{\dagger
		}(x,t)\psi _{1}(x,t)\biggr]}|\Phi (0)\rangle _{1} \\
	&\times &{}_{1}\langle \Phi (0)|T^{-1}\exp {\biggl[i\frac{2mc}{\hbar }\sqrt{%
			\frac{\pi G}{N-1}}\int dt\,dx[\varphi ^{\prime }(x,t)+\varphi _{1}^{\prime
		}(x,t)]\psi _{1}^{\dagger }(x,t)\psi _{1}(x,t)\biggr]|z_{k}\rangle _{1}}
\end{eqnarray*}

\begin{eqnarray*}
	&=&\int \mathcal{D}[\varphi ]\prod_{\alpha }\mathcal{D}[\varphi _{\alpha }]%
	\mathcal{D}[\varphi ^{\prime }]\prod_{\alpha }D[\varphi ^{\prime }{_{\alpha }%
	}]exp{\biggl[\frac{ic^{2}}{2\hbar }\int dt\,dx\bigl[\varphi \nabla
		^{2}\varphi -\sum_{\alpha }\varphi _{\alpha }\nabla ^{2}\varphi _{\alpha
		}-\varphi ^{\prime 2}\varphi ^{\prime }+\sum_{\alpha }\varphi _{\alpha
		}^{\prime }\nabla ^{2}\varphi _{\alpha }^{\prime }\bigr]}\biggr] \\
	&\times &\biggl[\bigotimes_{\alpha =2}^{N}\frac{1}{\sqrt{\mathcal{N}}}%
	\sum_{j=1}^{\mathcal{N}}\,{}_{\alpha }\langle z_{j}|\biggr]\exp {\biggl[i%
		\frac{2mc}{\hbar }\sqrt{\frac{\pi G}{N-1}}\sum_{\alpha =2}^{N}\int
		dt\,dx[\varphi ^{\prime }(x,t)+\varphi _{\alpha }^{\prime }(x,t)]n(x-z_{j})%
		\biggr]} \\
	&\times &\exp {\biggl[-i\frac{2mc}{\hbar }\sqrt{\frac{\pi G}{N-1}}%
		\sum_{\alpha =2}^{N}\int dt\,dx[\varphi (x,t)+\varphi _{\alpha
		}(x,t)]n(x-z_{j})\biggr]}\biggl[\bigotimes_{\alpha =2}^{N}\frac{1}{\sqrt{%
			\mathcal{N}}}\sum_{j=1}^{\mathcal{N}}|z_{j}\rangle _{\alpha }\biggr]\times \\
	&\times &\exp {\biggl[-i\frac{2mc}{\hbar }\sqrt{\frac{\pi G}{N-1}}\int
		dt\,dx[\varphi (x,t)+\varphi _{1}(x,t)]n(x-z_{h})\biggr]} \\
	&\times &\exp {\biggl[i\frac{2mc}{\hbar }\sqrt{\frac{\pi G}{N-1}}\int
		dt\,dx[\varphi ^{\prime }(x,t)+\varphi _{1}^{\prime }(x,t)]n(x-z_{k})\biggr]}%
	\times {}_{1}\langle z_{h}|\Phi (0)\rangle _{1}\langle \Phi (0)|z_{k}\rangle
	_{1}
\end{eqnarray*}%
\begin{eqnarray}
	&=&\frac{1}{\mathcal{N}^{N}}\sum_{j_{2},\dots ,j_{a}\dots ,j_{N}}\exp {%
		\biggl[-\frac{im^{2}}{2\hbar }\frac{G}{N-1}t\int dx\,dy\frac{[\sum_{\alpha
				=2}^{N}n(x-z_{j_{\alpha }})+n(x-z_{h})][\sum_{\alpha ^{\prime
				}=2}^{N}n(y-z_{j_{\alpha ^{\prime }}})+n(y-z_{h})]}{|x-y|}\biggr]}  \notag \\
	&\times &\exp {\biggl[\frac{im^{2}}{2\hbar }\frac{G}{N-1}t\int dx\,dy\frac{%
			[\sum_{\alpha =2}^{N}n(x-z_{j_{\alpha }})+n(x-z_{k})][\sum_{\alpha ^{\prime
				}=2}^{N}n(y-z_{j_{\alpha ^{\prime }}})+n(y-z_{k})]}{|x-y|}\biggr]}  \notag \\
	&\times &\exp {\biggl[-\frac{im^{2}}{2\hbar }\frac{G}{N-1}t\int dx\,dy\biggl[%
		\frac{n(x-z_{h})n(y-z_{h})+\sum_{\alpha =2}^{N}n(x-z_{j_{\alpha
			}})n(y-z_{j_{\alpha }})}{|x-y|}\biggr]\biggr]}  \notag \\
	&\times &\exp {\biggl[-\frac{im^{2}}{2\hbar }\frac{G}{N-1}t\int dx\,dy\biggl[%
		\frac{n(x-z_{k})n(y-z_{k})+\sum_{\alpha =2}^{N}n(x-z_{j_{\alpha
			}})n(y-z_{j_{\alpha }})}{|x-y|}\biggr]\biggr]}  \notag \\
	&=&\frac{1}{\mathcal{N}^{N}}\sum_{j_{2},\dots ,j_{a}\dots ,j_{N}}\exp {%
		\biggl[\frac{im^{2}}{2\hbar }\frac{G}{N-1}t\int dx\,dy\frac{[\sum_{\alpha
				=2}^{N}n(x-z_{j_{\alpha }})n(y-z_{h})-\sum_{\alpha =2}^{N}n(x-z_{j_{\alpha
			}})n(y-z_{k})]}{|x-y|}\biggr]}  \label{espo} \\
	&=&\frac{1}{\mathcal{N}^{N}}\biggl[\sum_{j=1}^{\mathcal{N}}e^{i\frac{%
			\mathcal{A}_{j}}{N-1}}\biggr]^{N-1}, \notag
\end{eqnarray}

where

\begin{equation*}
	\mathcal{A}_{j}={\frac{m^{2}G\ t}{2}\int dx\,dy\frac{%
			[n(x-z_{j})n(y-z_{h})-n(x-z_{j})n(y-z_{k})]}{|x-y|}.}
\end{equation*}

Now, using the property of the multinomial distributions for $Multi(n,%
\mathbf{p})\underset{n\rightarrow \infty }{\approx }\mathfrak{N}(n\mathbf{p}%
,n\Sigma ),$ where $\mathfrak{N}$ is the normal multivariate distribution
and $\Sigma =P-\mathbf{p}\mathbf{p}^{T},$ with $P$ being the diagonal matrix
whose diagonal is formed by the probability vector $\mathbf{p},$ we can rewrite (%
\ref{espo}) as
\begin{equation*}
	\frac{1}{\mathcal{N}^{N}}\sum_{\substack{ k1,\dots ,k_{\mathcal{N}} \\ %
			\mbox{with }k_{1}+\dots +k_{\mathcal{N}}=N-1}}\binom{N-1}{k_{1}k_{2}\dots k_{%
			\mathcal{N}}}\biggl[\frac{e^{i\frac{\mathcal{A}_{1}}{N-1}}}{\mathcal{N}}%
	\biggr]^{k_{1}}\biggl[\frac{e^{i\frac{\mathcal{A}_{2}}{N-1}}}{\mathcal{N}}%
	\biggr]^{k_{2}}\dots \biggl[\frac{e^{i\frac{\mathcal{A}_{\mathcal{N}}}{N-1}}%
	}{\mathcal{N}}\biggr]^{k_{\mathcal{N}}},
\end{equation*}%
which, introducing the rescaled variables $\chi _{i}\equiv \frac{k_{i}}{N-1}$
and passing from sum to integral, transforms into
\begin{eqnarray}
	&&\frac{e^{i\mathcal{A}_{\mathcal{N}}}}{\mathcal{N}}\int_{\substack{ \chi
			_{i}>0,\forall i=1,\dots ,\mathcal{N}-1 \\ \mbox{with}\sum_{i=1}^{\mathcal{N}%
				-1}\chi _{i}\leq 1}}d\chi _{1}\dots d\chi _{\mathcal{N}}\prod_{j=1}^{%
		\mathcal{N}-1}\frac{1}{\sqrt{2\pi }\sqrt{\frac{1}{\mathcal{N}(\mathcal{N}-1)}%
	}}\exp {\biggl[-\frac{(\chi _{j}-\frac{1}{\mathcal{N}})^{2}}{\frac{2}{%
				\mathcal{N}(N-1)}}+i(\mathcal{A}_{j}-\mathcal{A_{\mathcal{N}}})\chi _{j}%
		\biggr]}  \notag \\
	&\xrightarrow[N\rightarrow \infty]{}&\frac{e^{i\mathcal{A}_{\mathcal{N}}}}{%
		\mathcal{N}}\int_{\substack{ \chi _{i}>0,\forall i=1,\dots ,\mathcal{N}-1 \\ %
			\mbox{with}\sum_{i=1}^{\mathcal{N}-1}\chi _{i}\leq 1}}d\chi _{1}\dots d\chi
	_{\mathcal{N}}\delta \biggl(\chi _{1}-\frac{1}{\mathcal{N}}\biggr)\times
	\dots \times \delta \biggl(\chi _{\mathcal{N}-1}-\frac{1}{\mathcal{N}}\biggr)%
	e^{i(\mathcal{A}_{j}-\mathcal{A}_{\mathcal{N}})\chi _{j}}=  \notag \\
	&=&\frac{1}{\mathcal{N}}e^{i/\mathcal{N}\sum_{j=1}^{\mathcal{N}}\mathcal{A}%
		_{j}}.  \label{lunga}
\end{eqnarray}%
Then

\begin{equation*}
	\left\vert _{1}\langle z_{h}|M(t)|z_{k}\rangle _{1}\right\vert \underset{%
		N\rightarrow \infty }{\longrightarrow }\frac{1}{\mathcal{N}},
\end{equation*}

while, in the same limit, the spreading time tends to infinity, from which we
conclude that the mechanism of random phase cancelation, leading for finite $%
N$ to a rapid decoherence of the superposition of localized states on one
side, and the spreading of the wave function on the other side, are
completely suppressed in the limit $N\rightarrow \infty $.

It is worth noting that, for the simplest case of $N=2$ treated in the main
text, we get easily the expression for the characteristic state reduction
time. Starting from the intermediate expression \ref{espo} specialized
for $N=2$,

\begin{equation*}
	{}\frac{1}{\mathcal{N}^{2}}\sum_{j=1}^{\mathcal{N}}\exp {\biggl[\frac{
			im^{2}G\ t}{2\hbar }\int dx\,dy\frac{
			[n(x-z_{j})n(y-z_{h})-n(x-z_{j})n(y-z_{k})]}{|x-y|}\biggr],}
\end{equation*}

we see that phase cancelation occurs when exponentials reach values of order
$1$, which happens in a characteristic time $\tau _{g}=\hbar
G^{-1}M^{-5/3}\rho ^{-1/3}\ $ \cite{DeFMaimPRD}. For example, for a very fine
grain of sand, of mass $10^{-6}gr.$, we have $\tau _{g}\sim 10^{-10}\sec .$%
, consistently short with respect to the mass-independent spreading time of $%
10^{3}\sec .$

\end{appendices}

\end{document}